\documentclass[a4paper,12pt,english,german]{article}

\begin{document}

\centerline{\bf A limitation of the Nakajima-Zwanzig projection
method\footnote{Project financially supported by Ministry of
Science Serbia contract no 171028 and for MD also partially
supported by the ICTP - SEENET-MTP grant PRJ-09 (Cosmology and
Strings) within the framework of the SEENET-MTP Network.}}

\bigskip

M. Arsenijevi\' c$^a$\footnote{Corresponding author. Email:
momirarsenijevic@gmail.com}, J. Jekni\' c-Dugi\' c$^b$, M. Dugi\'
c$^a$

\smallskip

$^a${Department of Physics, Faculty of Science, 34 000 Kragujevac,
Serbia}

$^b${Department of Physics, Faculty of Science and Mathematics, 18
000 Ni\v s, Serbia}

\bigskip

{\bf Abstract.} There is current interest in dynamical description
of different decompositions of a quantum system into subsystems.
We investigate usefulness of the Nakajima-Zwanzig projection
method in this context. Particularly, we are interested in
simultaneous description of dynamics of open systems pertaining to
different system-environment splits (decompositions). We find that
the Nakajima-Zwanzig and related projection methods are
system-environment split specific, that every system-environment
split requires specific projector, and that projector adapted to a
split neither provides information about nor commute with a
projector adapted to an alternative system-environment split. Our
findings refer to finite- and infinite-dimensional systems and to
arbitrary kinds of system-environment splitting. These findings
are a direct consequence of the recently established quantum
correlations relativity. We emphasize the subtlety and delicacy
required of the task of simultaneously describing the dynamics of
alternate system-environment splits.

\bigskip


\bigskip

{\bf 1. Introduction}

\bigskip

There is current interest in dynamics of different decompositions
of a composite quantum system into subsystems$^{[1-13]}$. While
physical motivations, methods and technical details are diverse,
there is a common core of the task that can be investigated on the
sufficiently general background. Typically, a $C$ system is
decomposed as $C=S+E$. The point is that there are many possible
such decompositions, $S+E$, $S'+E'$, etc., and one is interested
e.g. in (a) dynamics of some subsystems, e.g. $S$, $S'$,  as well
is in (b) amount and dynamics of non-classical correlations
present in different decompositions.

Within the decoherence theory,$^{[14,15,16,17]}$ the task is
foundational. There is not  {\it a priori} privileged
system-environment split (decomposition)$^{[6,7,14,15,16,17]}$.
Quantum decoherence is typically studied starting from a fairly
unprincipled choice of system-environment split. In this sense,
decoherence is often considered as an approximate
process$^{[15,17]}$. In practice, and particularly for the
"macroscopic" world, division of the total system into "(open)
system" and "environment" is "imperfect"$^{[15]}$. In the quantum
optics context,$^{[4]}$ the task tackles validity of certain kind
of master equations. In quantum electrodynamics, the task points
out important implications for the ontology of noncovariant
canonical QED due to the gauge freedom$^{[5]}$. In quantum
information science, one can show that amount of quantum
(non-classical) correlations is not merely a feature of a
composite system, or of the system's state, but is a feature of
the composite system's split into subsystems$^{[8]}$. The task
equally refers to the system's split into "virtual subsystems"
with the question of existence of the preferable split (structure)
of the composite system$^{[7,9]}$. Different partitions [not
necessarily bipartitions] into subsystems, where some subsystems
may serve as the open system and the rest as the environment, is
perhaps the main model considered in the quantum phase transitions
field$^{[10]}$ (and the references therein). Tensor products of an
algebra of operators or observables of a composite system is the
main formal framework for the task regarding the
finite-dimensional systems$^{[1,2,3,10,11]}$. In quantum
information practice, one can hope to be able to recognize "the
preferable behavior of quantum correlations which allows a given
quantum system to be more flexible in applications."$^{[9]}$. So,
the task is of both academic as well as of interest for different
kinds of applications. A large interest in the topic regarding
applications can be found e.g. in Ref. [11]. Some prospects can be
found in Refs. [12,13]. For a recent review of a part of the topic
see Ref. [13].

In this paper, we launch a variant of the task that is of the kind
(a) described above and that is a common core for the most of the
research results$^{[1-13]}$. Actually, we are interested in {\it
simultaneous} dynamics of a pair of {\it open} systems formally
denoted $S$ and $S'$, which pertain to different decompositions
(structures) of a composite system. Specific for our
considerations is that we consider the task in the context of the
Nakajima-Zwanzig projection method in the open quantum systems
theory$^{[18,19]}$.

A part of motivation for the present paper is the fact that the
Nakajima-Zwanzig projection method$^{[20,21]}$ is {\it central} to
modern open quantum systems theory$^{[18,19]}$. The method
provides a systematic theoretical approach to Markovian dynamics
and sets a basis for a systematic (perturbative) approach to
non-Markovian dynamics$^{[18,19]}$. The open systems theory
provides the foundations of quantum measurement theory$^{[3,22]}$,
decoherence$^{[1,2,3]}$, and the emergence of thermodynamic
behavior$^{[23]}$. Applications of the theory of open quantum
systems are found in practically all areas of physics, ranging
from quantum optics$^{[24]}$, quantum information$^{[25]}$ and
condensed-matter physics$^{[26]}$ to chemical physics$^{[27]}$ and
spintronics$^{[28]}$.

The key idea behind the Nakajima-Zwanzig projection method
consists of the introduction of a certain projection operator,
$\mathcal{P}$, which acts on the operators of the state space of
the total system "system+environment" ($S+E$). If $\rho$ is the
density matrix of the total system, the projection
$\mathcal{P}\rho$ (the "relevant part" of the total density
matrix) serves to represent a simplified effective description
through a reduced state of the total system. The complementary
part (the "irrelevant part" of the total density matrix),
$\mathcal{Q} \rho = (I - \mathcal{P})\rho$. For the "relevant
part", $\mathcal{P}\rho(t)$, one derives closed equations of
motion in the form of integro-differential equation. The open
system's density matrix $\rho_S(t) = tr_E \mathcal{P}\rho(t)$
contains {\it all} necessary information about the open system
$S$.

The Nakajima-Zwanzig projection method assumes a concrete, in
advance chosen and fixed, system-environment split (a
"structure"), $S+E$, which is uniquely defined by the associated
tensor product structure (TPS) of the total system's Hilbert
space, $\mathcal{H} = \mathcal{H}_S \otimes \mathcal{H}_E$.
Division of the composite system into "system" and "environment"
is practically motivated. In principle, the projection method can
equally describe arbitrary system-environment split i.e. arbitrary
factorization of the total system's Hilbert state. At the time
when importance of quantum correlations was not acknowledged, the
Nakajima-Zwanzig method appeared to be a whole that can not and
should not be improved. But the existence of non-classical
correlations shed new light on versatility of  the method.

Interestingly enough, we find that the projection methods are
generally {\it unsuitable} for the task of {\it simultaneous}
description of  {\it open} systems $S$ and $S'$. Our finding is
general: it refers to the finite- as well as the
infinite-dimensional systems and to all kinds of the variables
transformations, which induce the tensor-product structures of the
composite system's Hilbert state space. Our results are due
ultimately to the recently established quantum correlations
relativity$^{[8]}$. It is therefore not surprising that we are
only now able to distinguish the following findings as the basis
for our main result: (i) first, [not very surprisingly], every
system-environment split requires a specific projector; (ii) the
projection-based information about the $S$ system is in general
not sufficient for drawing information about the $S'$ system at
the {\it same} time; (iii) one cannot construct mutually
compatible (commuting) projectors that pertain to different
decompositions {\it simultaneously}.

Our findings, that go beyond the  standard thinking$^{[18,19]}$ of
the open quantum systems, do not present any inconsistency with
the open systems theory or with the foundations of the
Nakajima-Zwanzig method. Rather, our findings point out that the
Nakajima-Zwanzig projection method has a {\it limitation}, i.e. is
{\it not suitable} for the above posed task. Finally, we emphasize
subtlety and delicacy of simultaneous dynamical description of the
open systems pertaining to  different system-environment splits.

\bigskip

{\bf 2. Simultaneous dynamics of the structures}

\bigskip

A composite system $C$ can be differently decomposed into
"system+environment", $S+E$ and $S'+E'$. We wonder if, within the
projection method, the unitary dynamics of $C$ can provide {\it
simultaneous} (i.e. in the same time interval) reduced dynamics
for both open systems, $S$ and $S'$. While description of the
different structures at the same time is basic--notice the
simultaneous redefinition of both "system" and "environment"--it
can also have some {\it practical} motivations. E.g. we can wonder
if the $S'$ system can be more easily accessible in a laboratory
than the $S$ system. This can provide the more convenient recipes
for manipulating the $C$'s degrees of freedom. Or we may be
interested in dynamics of the $S'$ system, which is not directly
accessible in a laboratory.

Quantum mechanics is insensitive to different structures
(decompositions in to parts) of a composite system $C$. That is,
quantum mechanics equally treats the different structures of  $C$.
The von Neumann-Liouville equation ($\hbar = 1$):

\begin{equation}
\label{eq.1}
 \frac{d\rho(t)}{ dt} = - \imath [H, \rho(t)]
\end{equation}

\noindent where $H$ is the total system's Hamiltonian acting on
the total system's Hilbert state-space $\mathcal{H}$, equally
applies to every decomposition (structure) of $C$.

Consider a pair of structures, $S+E$ and $S'+E'$; $S+E = C =
S'+E'$. This provides the different tensor-product-structures for
the total Hilbert state space, $\mathcal{H}_S \otimes
\mathcal{H}_E = \mathcal{H} = \mathcal{H}_{S'} \otimes
\mathcal{H}_{E'}$. Of course, the total system's Hamiltonian and
state, as well as reduced state of any subsystem (obtained by the
proper tracing out) are unique in every instant of time. The
different forms of the total Hamiltonian

\begin{equation}
\label{eq.2}
 H^{(SE)} \equiv H_S + H_E + H_{SE}\\ = H =  H_{S'} +
H_{E'} + H_{S'E'} \equiv H^{(S'E')},
\end{equation}

\noindent where the double subscripts distinguish the interaction
terms. Then eqs. (\ref{eq.1}) and (\ref{eq.2}) provide
simultaneous description of the reduced dynamics for both open
systems, $S$ and $S'$:

\begin{equation}
\label{eq.3}
 {d\rho_i(t) \over dt} = - \imath tr_j [H, \rho(t)],
\quad i=S,S', \quad j = E, E';
\end{equation}

\noindent for the $i=S$, the Hamiltonian $H$ takes the form
$H^{(SE)}$, while for the $i=S'$, the Hamiltonian takes the form
$H^{(S'E')}$.

The technical difficulties in solving  equations eq.(\ref{eq.3})
have historically led to the development of  the different
methods, notably to the Nakajima-Zwanzig projection
method$^{[20,21]}$, which introduces a projection operator
$\mathcal{P}$ and its complementary projection operator,
$\mathcal{Q} = I - \mathcal{P}$. The projection $\mathcal{P} \rho
(t)$ is {\it required} to contain {\it all} necessary information
about the open system $S$:

\begin{equation}
\label{eq.4}
 \rho_S(t) = tr_E \rho(t) = tr_E \mathcal{P} \rho(t)
 \Leftrightarrow tr_E \mathcal{Q}\rho(t) = 0,
\forall{t}.
\end{equation}

Then the task is to provide a closed master equation for
$\rho_S(t)$, such as e.g. the generalized Nakajima-Zwanzig
equation or the time-convolutionless  master equation$^{[18,19]}$.

The linear projections fulfilling eq.(\ref{eq.4}) can be
defined$^{[14,23]}$: (i) $\mathcal{P}\rho(t) = (tr_E \rho(t))
\otimes \rho_E$ [for some $\rho_E \neq tr_S \rho$], (ii)
$\mathcal{P} \rho(t) = \sum_n (tr_E P_{Sn} \rho(t)) \otimes
\rho_{En}$ [with arbitrary orthogonal supports for $\rho_E$s], and
(iii) $\mathcal{P} \rho(t) = \sum_i (tr_E P_{Ei} \rho(t)) \otimes
P_{Ei}$ [with arbitrary orthogonal projectors for the $E$ system];
by $P$, we denote the projectors on the respective Hilbert state
(factor) spaces. The physical context fixes the choice of the
projection--e.g. by an assumption about the initial state. In this
paper we stick to the projection (i), which is by far of the
largest interest in foundations and applications of the open
systems theory.

To illustrate our point, consider the standard quantum
teleportation setup [25] for three qubits, $C = 1 + 2 +3$. The two
bipartite structures, $S+E \equiv 1 + (2+3)$ and $S'+E' \equiv
(1+2) + 3$. Then the quantum state $\vert \Psi \rangle$ of $C$ can
be written as [25]

\begin{equation}
\label{eq.5}
 \vert u \rangle_{S} \otimes \vert \phi\rangle_{E} =
\vert \Psi\rangle = \sum_{i=1}^4 {1 \over 2} \vert i \rangle_{S'}
\otimes \vert i \rangle_{E'} .
\end{equation}

The projection (i) gives $\mathcal{P'} \rho = (1/4) \sum_i \vert i
\rangle_{S'}\langle i \vert \otimes \rho_{E'}$ for the $S'+E'$
structure that is a mixed state fulfilling the condition
eq.(\ref{eq.4}). Application of the same procedure for the $S+E$
structure, i.e. the projection (i), gives the equality
$\mathcal{P} \rho = \rho$ for the $S+E$ structure also satisfying
eq.(\ref{eq.4}). The projector $\mathcal{P}'$ provides a mixed
state for the total system. However, the $\mathcal{P}$ projector
provides a pure state for the total system. So, the two projection
operators, $\mathcal{P}$ and $\mathcal{P}'$, cannot equal to each
other. Moreover, as they provide different states e.g. for the $1$
subsystem in an instant in time, the two projectors {\it exclude}
each other.  The same conclusion applies to the hydrogen atom
differently structured either as "electron+proton ($e+p$)" or as
the "center of mass + relative position ($CM+R$)". The atom's
(instantaneous) state $\vert \Phi\rangle$$^{[17,29]}$

\begin{equation}
\label{eq.6}
 \sum_l c_l \vert l \rangle_e \otimes \vert l
\rangle_p = \vert \Phi \rangle = \vert \chi \rangle_{CM} \otimes
\vert n l m_l m_s \rangle_R;
\end{equation}

\noindent in eq.(\ref{eq.6}), the quantum numbers $n, l, m_l, m_s$
are the standard quantum numbers known from the quantum theory of
the hydrogen atom.

The equalities eqs.(\ref{eq.5}) and (\ref{eq.6}) are instances of
a quantum mechanical rule  of "entanglement relativity"
(ER),$^{[1,2,3,6,7,8,11,12,13]}$ (and the references therein),
which has recently been extended to relativity of the more general
non-classical (quantum) correlations quantified by
"discord"$^{[8]}$--quantum correlations relativity
(QCR)$^{[8,13]}$. Quantum correlations relativity emphasizes: a
transformation of (a change in) the degrees-of-freedom typically
effects in a {\it change} in correlations present in the composite
system $C$. For instance (for arbitrary instant in time), a tensor
product state for one structure is endowed by non-classical
correlations for the alternate structure--e.g. for a mixed state
$\rho$ ($\rho^2 \neq \rho, tr \rho = 1$)$^{[8,13]}$

\begin{equation}
\label{eq.7}
 \rho_S \otimes \rho_E = \rho = \sum_i \lambda_i
\rho_{S'i} \otimes \rho_{E'i}, \quad \sum_i \lambda_i = 1,
\end{equation}

\noindent where in general the density matrices for the $S'+E'$
structure are neither linearly dependent nor commuting. Exceptions
to QCR are not ruled out. However, such exceptions become
 irrelevant in the dynamical analysis to be presented below.

If eq.(\ref{eq.7}) is a consequence of the projection (i) (when
the $\rho$ in eq.(\ref{eq.7}) should be exchanged by
$\mathcal{P}\rho$) then it contains all  necessary information
about the open system $S$, i.e. the requirement eq.(\ref{eq.4}) is
fulfiled. However, the rhs of eq.(\ref{eq.7}) is in general not of
any type of the above distinguished projections (i)-(iii).
Likewise for eqs.(\ref{eq.5}) and (\ref{eq.6}), the projection
$\sum_i \lambda_i \rho_{S'i} \otimes \rho_{E'i}$ does not in
general encapsulate all necessary information about the open
system $S'$. Rather, the "irrelevant part", $\mathcal{Q}\rho$, can
be expected to bring some information about the open system $S'$.

Given eq.(\ref{eq.4}) is fulfilled, eq.(\ref{eq.7}) (with the
$\mathcal{P}\rho$ instead of $\rho$) implies:

\begin{equation}
\label{eq.8}
 tr_E \mathcal{Q}\rho(t) = tr_E (\rho(t) -
\mathcal{P}\rho(t)) =  tr_E (\rho(t) - \rho_S(t) \otimes \rho_E) =
0, \forall{t}.
\end{equation}

The analog condition regarding the $S'+E'$ structure, for the same
time instant:

\begin{equation}
\label{eq.9}
 tr_{E'} \mathcal{Q}\rho(t) = tr_{E'} (\rho(t) -
\rho_S(t) \otimes \rho_E) = 0, \forall{t}.
\end{equation}

Then both $tr_E \mathcal{P}\rho(t) = \rho_S(t)$ and $tr_{E'}
\mathcal{P}\rho(t) = \rho_{S'}(t)$ and one can write sumultaneos
(for the same time interval) master equations for the $S$ and $S'$
systems, with the constraints coming from eq.(\ref{eq.7}).

However, eq.(\ref{eq.9}) is  {\it not} fulfilled. More precisely:

\noindent
 {\bf Lemma 1.} For the most part of the composite system's
dynamics, validity of eq.(\ref{eq.8}) implies nonvalidity of
eq.(\ref{eq.9}), and {\it vice versa}.

Proof: Given eq.(\ref{eq.8}), i.e. $tr_E \mathcal{Q}\rho(t) = 0,
\forall{t}$, we investigate the conditions that should be
fulfilled in order for eq.(\ref{eq.9}), i.e. $tr_{E'}
\mathcal{Q}\rho(t) = 0, \forall{t}$, to be fulfilled. The
$\mathcal{Q}$ projector refers to the $S+E$, not to the $S'+E'$
structure. Therefore, in order to calculate
$tr_{E'}\mathcal{Q}\rho(t)$, we use ER. We refer to the projection
(i) in an instant of time:

\begin{equation}
\label{eq.10}
 \mathcal{P} \rho = (tr_E \rho) \otimes \rho_E.
\end{equation}

\noindent A) Pure state $\rho = \vert \Psi \rangle\langle \Psi
\vert$, while, due to eq.(\ref{eq.8}), $tr_E \mathcal{Q} \vert
\Psi \rangle\langle \Psi \vert = 0$.

We consider the pure state presented in its (not necessarily
unique) Schmidt form

\begin{equation}
\label{eq.11}
 \vert \Psi \rangle = \sum_i c_i \vert i\rangle_S
\vert i\rangle_E,
\end{equation}

\noindent where $\rho_S = tr_E \vert \Psi\rangle\langle \Psi\vert
= \sum_i p_i \vert i\rangle_S\langle i\vert$, $p_i = \vert
c_i\vert^2$ and for arbitrary $\rho_E \neq tr_S \vert
\Psi\rangle\langle\Psi\rangle$. Given $\rho_E = \sum_{\alpha}
\pi_{\alpha} \vert \alpha \rangle_E\langle \alpha \vert $, we
decompose $\vert \Psi\rangle$ as:

\begin{equation}
\label{eq.12} \vert \Psi\rangle = \sum_{i, \alpha} c_i C_{i\alpha}
\vert i\rangle_S \vert \alpha\rangle_E,
\end{equation}

\noindent with the constraints:

\begin{equation}
\label{eq.13}
 \sum_i \vert c_i\vert^2 = 1 = \sum_{\alpha}
\pi_{\alpha}, \sum_{\alpha} \vert C_{i\alpha} \vert^2 = 1,
\forall{i},
\end{equation}

Then

\begin{equation}
\label{eq.14}
 \mathcal{Q} \vert \Psi\rangle\langle \Psi \vert =
\vert \Psi \rangle\langle \Psi \vert - \sum_{i, \alpha} p_i
\pi_{\alpha} \vert i\rangle_S\langle i\vert \otimes \vert \alpha
\rangle_E\langle \alpha \vert.
\end{equation}

We use ER:

\begin{equation}
\label{eq.15}
 \vert i\rangle_S \vert \alpha\rangle_E = \sum_{m,n}
D^{i\alpha}_{mn} \vert m\rangle_{S'} \vert n \rangle_{E'}
\end{equation}

\noindent with the constraints:

\begin{equation}
\label{eq.16}
 \sum_{m,n}  D^{i\alpha}_{mn} D^{i'\alpha'\ast}_{mn}
= \delta_{ii'} \delta_{\alpha\alpha'}.
\end{equation}

With the use of eqs.(\ref{eq.12}) and (\ref{eq.15}),
eq.(\ref{eq.14}) reads:

\begin{equation}
\label{eq.17}
 \sum_{m,m'n,n'}[\sum_{i,i', \alpha, \alpha'} c_i
C_{i\alpha} c_{i'}^{\ast}C_{i'\alpha'}^{\ast} D^{i\alpha}_{mn}
D^{i'\alpha' \ast}_{m'n'} - \sum_{i,\alpha} p_i \pi_{\alpha}
D^{i\alpha}_{mn} D^{i\alpha\ \ast}_{m'n'}] \vert
m\rangle_{S'}\langle m' \vert \otimes \vert n\rangle_{E'}\langle
n'\vert.
\end{equation}

After tracing out, $tr_{E'}$:

\begin{equation}
\label{eq.18}
 \sum_{m,m'} \left.\{  \sum_{i,\alpha,n}
\sum_{i',\alpha'} c_i C_{i\alpha}
c_{i'}^{\ast}C_{i'\alpha'}^{\ast} D^{i\alpha}_{mn}
D^{i'\alpha'\ast}_{m'n}\right. -  \left. p_i \pi_{\alpha}
D^{i\alpha}_{mn} D^{i\alpha\ast}_{m'n} \right.\} \vert
m\rangle_{S'}\langle m'\vert
\end{equation}

Hence

\begin{equation}
\label{eq.19}
 tr_{E'} \mathcal{Q} \vert \Psi\rangle\langle
\Psi\vert = 0 \Leftrightarrow \sum_{i,\alpha,n} [\sum_{i',\alpha'}
c_i C_{i\alpha} c_{i'}^{\ast} C_{i'\alpha'}^{\ast}
D^{i\alpha}_{mn} D^{i'\alpha'\ast}_{m'n} -  p_i \pi_{\alpha}
D^{i\alpha}_{mn} D^{i\alpha\ast}_{m'n}] = 0, \forall{m,m'}.
\end{equation}

Introducing notation, $\Lambda^m_n \equiv \sum_{i,\alpha} c_i
C_{i\alpha} D^{i\alpha}_{mn}$, one obtains:

\begin{equation}
\label{eq.20}
 tr_{E'} \mathcal{Q} \vert \Psi\rangle\langle
\Psi\vert = 0 \Leftrightarrow \\  A_{mm'} \equiv \sum_{n}
[\Lambda^m_n \Lambda^{m'\ast}_n - \sum_{i,\alpha}p_i \pi_{\alpha}
D^{i\alpha}_{mn} D^{i\alpha\ast}_{m'n}] = 0, \forall{m,m'}.
\end{equation}

Notice:

\begin{equation}
\label{eq.21} \sum_m A_{mm} = 0.
\end{equation}

\noindent which is equivalent to $tr \mathcal{Q} \vert \Psi\rangle
\langle \Psi \vert = 0$, see eq.(\ref{eq.14}).

\noindent B) Mixed (e.g. non-entangled) state.

\begin{equation}
\label{eq.22}
 \rho = \sum_i \lambda_i \rho_{Si}\rho_{Ei}, \quad
\rho_{Si} = \sum_m p_{im} \vert \chi_{im}\rangle_S
\langle\chi_{im} \vert,   \\ \rho_{Ei} = \sum_n \pi_{in}\vert
\phi_{in}\rangle_E\langle \phi_{in}\vert,
\end{equation}

In eq.(\ref{eq.22}), having in mind eq.(\ref{eq.10}), $tr_E
\mathcal{Q} \rho = 0$, while $tr_E\rho = \sum_p \kappa_p \vert
\varphi_{p}\rangle_S\langle \varphi_{p}\vert$, and $\rho_E =
\sum_q \omega_q \vert \psi_{q}\rangle_E\langle \psi_{q}\vert \neq
tr_S \rho$.

Constraints:

\begin{equation}
\label{eq.23}
 \sum_i \lambda_i = 1 = \sum_p \kappa_p = \sum_q
\omega_q, \quad \sum_{m} p_{im} = 1 = \sum_n \pi_{in}, \forall{i}.
\end{equation}

Now we make use of ER and, for comparison, we use the same basis
$\{\vert a\rangle_{S'} \vert b \rangle_{E'}\}$

\begin{equation}
\label{eq.24}
 \vert \chi_{im}\rangle_S \vert \phi_{in} \rangle_E =
\sum_{a,b} C^{imn}_{ab} \vert a \rangle_{S'} \vert b \rangle_{E'},
 \\ \vert\varphi_p\rangle_S \vert \psi_q\rangle_E = \sum_{a,b}
D^{pq}_{ab} \vert a \rangle_{S'} \vert b \rangle_{E}.
\end{equation}

Constraints:

\begin{equation}
\label{eq.25}
 \sum_{a,b} C^{imn}_{ab} C^{im'n'\ast}_{ab} =
\delta_{mm'} \delta_{nn'}, \quad \sum_{a,b} D^{pq}_{ab}
D^{p'q'\ast}_{ab} = \delta_{pp'} \delta_{qq'}.
\end{equation}

So

\begin{equation}
\label{eq.26} \mathcal{Q}\rho = \rho - (tr_E\rho) \otimes \rho_E =
\sum_{a,a',b,b'} \{\sum_{i,m,n} \lambda_i p_{im} \pi_{in}
C^{imn}_{ab} C^{imn\ast}_{a'b'}  \\ - \sum_{p,q} \kappa_p \omega_q
D^{pq}_{ab} D^{pq\ast}_{a'b'}\} \vert a \rangle_{S'}\langle
a'\vert \otimes \vert b\rangle_{E'} \langle b'\vert.
\end{equation}

Hence

\begin{equation}
\label{eq.27}
 tr_{E'} \mathcal{Q} \rho = 0 \Leftrightarrow
\Lambda_{aa'} \equiv \sum_{i,m,n,b} \lambda_i p_{im} \pi_{in}
C^{imn}_{ab} C^{imn\ast}_{a'b}-  \sum_{p,q,b} \kappa_p \omega_q
D^{pq}_{ab} D^{pq\ast}_{a'b} = 0, \forall{a,a'}.
\end{equation}

Again, for $a=a'$:

\begin{equation}
\label{eq.28}
 \sum_a \Lambda_{aa} = 0,
\end{equation}

\noindent as being equivalent with $tr \mathcal{Q} \rho = 0$, see
eq.(\ref{eq.26}).

Validity of eq.(\ref{eq.9}) assumes validity of eq.(\ref{eq.20})
for pure and of eq.(\ref{eq.27}) for mixed states. Both
eq.(\ref{eq.20}) and eq.(\ref{eq.27}) represent the sets of
simultaneously satisfied equations. We do not claim non-existence
of the particular solutions to eq.(\ref{eq.20}) and/or to
eq.(\ref{eq.27}), e.g. for the finite-dimensional systems.
Nevertheless, we want to emphasize that the number of states they
might refer to is apparently negligible compared to the number of
states for which this is not the case. For instance, already for
the fixed $a$ and $a'$, a small change e.g. in $\kappa$s (while
bearing eq.(\ref{eq.23}) in mind) undermines equality in
eq.(\ref{eq.27}).

Quantum dynamics is continuous in time. Provided eq.(\ref{eq.8})
is fulfilled, validity of eq.(\ref{eq.9}) might refer {\it only}
to a special set of the time instants. So we conclude: for the
most part of the open $S'$-system's dynamics, eq.(\ref{eq.9}) is
not fulfilled. By exchanging the roles of eq.(\ref{eq.8}) and
eq.(\ref{eq.9}) in our analysis, we obtain the reverse conclusion,
which completes the proof. Q.E.D.

Lemma 1 establishes: as long as eq.(\ref{eq.4}) (i.e.
eq.(\ref{eq.8})) is valid for every instant in time, the analogous
equality

\begin{equation}
\label{eq.29}
 \rho_{S'}(t) = tr_{E'} \rho(t) = tr_{E'} \mathcal{P}
\rho(t),
\end{equation}

\noindent cannot be fulfilled for the most part of the open
$S'$-system's dynamics, and {\it vice versa}. Then, as emphasized
above, for the most part of the composite system's dynamics, the
projection $\mathcal{Q}\rho$ ($\mathcal{Q}'\rho$) brings some
information about the open system $S'$ ($S$)--in {\it
contradiction} with the basic {\it idea} of the Nakajima-Zwanzig
projection method$^{[18,19,20,21]}$.

Regarding the simultaneous projecting:

\noindent {\bf Lemma 2.} The two structure-adapted projectors
$\mathcal{P}$ and $\mathcal{P}'$  do not mutually commute for the
projection  (i).

Proof: The commutation condition, $[\mathcal{P},
\mathcal{P}']\rho(t) = 0, \forall{t}$. With the notation
$\rho_P(t) \equiv \mathcal{P}\rho(t)$ and $\rho_{P'}(t) \equiv
\mathcal{P}'\rho(t)$, the commutativity reads:
$\mathcal{P}\rho_{P'}(t) = \mathcal{P}'\rho_P(t), \forall{t}$.
Then $\mathcal{P}\rho_{P'}(t) = tr_E\rho_{P'}(t) \otimes \rho_E =
\rho_S(t) \otimes \rho_E$, while $\mathcal{P}'\rho_{P}(t) =
tr_{E'}\rho_P(t) \otimes  = \sigma_{S'}(t) \otimes \sigma_{E'}$.
So, the commutativity requires the equality $\sigma_{S'}(t)\otimes
\sigma_{E'} = \rho_{S}(t)\otimes \rho_{E}, \forall{t}$. However,
quantum dynamics is continuous in time. Likewise in Proof of Lemma
1, quantum correlations relativity guarantees, that for the most
of the time instants the equality will not be fulfilled. Q.E.D.

Lemma 2 establishes: for any pair of structures, $S+E$ and
$S'+E'$, one {\it cannot} choose/construct a pair of compatible
projectors pertaining to the projection (i) in the same time
itnerval.

Thus the Nakajima-Zwanzig projection method faces a limitation.
While it can be separately performed for any structure (either
$\mathcal{P}$ or $\mathcal{P}'$), it cannot be {\it
simultaneously} used for a pair of structures. Once performed, the
projection does not in general allow for drawing complete
information about an alternative structure of the composite
system--projecting is non-invertible ("irreversible").

Our finding refers to {\it all projection-based methods}. In
formal terms: Lemma 1 implies that $d\mathcal{P}\rho(t)/dt$ allows
tracing out over only one structure of the composite system. If
that structure is $S+E$, then $tr_{E'} d\mathcal{P} \rho(t)/dt
\neq d\rho_{S'}(t)/dt$ [as long as $\rho_{S'}(t) =
tr_{E'}\rho(t)$]. That is, eq.(\ref{eq.4}) for the $S'+E'$
structure  is not fulfilled, and therefore cannot provide a
projection-based master equation for the $S'$ system. This can be
seen also from the following argument, which is not restricted to
the projection-based methods. Tracing out over $E$ is dependent
on, but not equal to, the tracing out over $E'$, and {\it vice
versa}. This dependence follows from the fact that the $S$ and $E$
degrees of freedom are intertwined with the $S'$ and $E'$ degrees
of freedom. Intuitively: "$tr_E$" (e.g. integrating over the $E$'s
degrees of freedom) partly encompasses both the $S'$ and the $E'$
degrees of freedom. On the other hand, Lemma 2 excludes
simultaneous projecting, i.e. simultaneous master equations for
the two structures. E.g., $d\mathcal{P}\rho(t)/dt = d\rho_S(t)/dt
\otimes \rho_E$ is in conflict with $d\mathcal{P}'\rho(t)/dt =
d\rho_{S'}(t)/dt \otimes \rho_{E'}$: due to eq.(\ref{eq.7}), only
one of them can be correct for arbitrary instant in time.

Summarizing: if we consider simultaneous (i.e. for the same time
interval) dynamics for the open systems pertaining to a pair of
system-environment splits, Lemma 1 establishes that projection
adapted to one structure cannot be used for deriving master
equation regarding another structure, while Lemma 2 emphasizes
that simultaneous projecting for the two structures is not
allowed. Hence, in order to simultaneously describe dynamics of
the two open systems, $S$ and $S'$, one should avoid projecting of
the composite system's state.

Our findings take us back to the beginning, i.e. to
eq.(\ref{eq.3}), which does not have any limitation. It seems that
there is not a universal shortcut in deriving master equations
regarding the alternative structures of a composite system.

\bigskip

{\bf 3. Discussion}

\bigskip

We are interested in the variables transformations that
simultaneously redefine both the open system and its environment.
The transformations include regrouping of the constituent
particles--e.g. in "entanglement swapping"$^{[25,30]}$, which is
illustrated by eq.(\ref{eq.5})--or the more general
transformations as illustrated by eq.(\ref{eq.6}). Such
transformations are examples of the more general linear canonical
transformations performed on the total system
"system+environment"$^{[6,9,10,13]}$. The transformations that
target only the open system without altering the environment can
be found in$^{[4,5,7]}$.

Investigating the alternative system-environment splits goes
beyond the standard, practically inspired methods in the open
systems theory. This new research line is still in its infancy but
is of interest for both academic as well as for applied research
(see Introduction for the references).

Despite the fact that QCR can have exceptions for certain states,
our findings presented by Lemma 1 and 2 do not. Even if QCR does
not apply to an instant in time (i.e. to a special state of the
total system), it is most likely to apply already for the next
instant of time in the unitary (continuous in time) dynamics of
the total system $C$. This general argument makes Lemma 1 and 2
universal, i.e. applicable to every Hilbert state space and every
model and structure (the choice of the open systems $S$ and $S'$)
of the total system. So Lemma 1 and Lemma 2 refer to the finite-
and infinite-dimensional systems and to all kinds of the variables
transformations.

Lemma 2 forbids construction of compatible projectors  for a pair
of the system-environment splits. So the only way to use the
projection method is to have fulfilled the conditions $\rho_S(t) =
tr_E \mathcal{P}\rho(t)$ and $\rho_{S'}(t) = tr_{E'}
\mathcal{P}\rho(t)$ for every instant in time $t$. However,
according to Lemma 1, these equalities (see eq.(\ref{eq.8}) and
eq.(\ref{eq.9})) cannot both be fulfilled for every instant in
time--moreover, the equalities are not fulfilled for the most of
the time instants.

Thus we are forced to conclude that the task of simultaneous
description of the different structures  reduces to
eq.(\ref{eq.3}), yet with the constraint imposed by  quantum
correlations relativity$^{[8,13]}$.

Consider the simplest case for the $S+E$ structure: the tensor
product initial state, the environment $E$ is harmonic bath of
non-interacting oscillators weakly interacting with the open
system $S$, applicability of the Born-Markov and of the
rotating-wave approximation. Then from eq.(\ref{eq.7}) we can
directly draw the following conclusions regarding the alternative
structure $S'+E'$: (a) due to the presence of the initial
correlations in eq.(\ref{eq.7}), the $S'$ system's dynamics
(described by eq.(\ref{eq.3})) is non-Markovian, and also possibly
non-completely positive$^{[18,19,31,32,33,34,35]}$; (b) the new
environment $E'$ is in general not in thermal equilibrium--in
general it is in non-stationary state; (c) in general, the
environment $E'$ consists of mutually interacting particles. In
addition to this, both the strength of interaction and validity of
the rotating wave approximation  can be at stake for the
alternative $S'+E'$ structure. Thus, in general, investigating the
alternative open system's dynamics is a formidable task, but
see$^{[6,9]}$.

\bigskip

{\bf 4. Conclusion}

\bigskip

Relativity, i.e. structure dependence, of quantum correlations
limits application of the Nakajima-Zwanzig and of the related
projection methods in investigating the system-environment splits.
The projection-methods-provided information about a subsystem of a
composite system in an instant of time is insufficient to acquire
information about another subsystem of the same composite system
in the same instant of time.  This limitation of the projection
methods suggests that  "shortcuts" for describing the alternative
system-environment-splitting dynamics may be non-reliable and
delicate.

\bigskip

{\bf References}

\bigskip

[1] Zanardi P 2001 {\it Phys. Rev. Lett.} {\bf 87} 077901

[2] Zanardi P, Lidar D A and Lloyd S 2004 {\it Phys. Rev. Lett.}
{\bf 92} 060402

[3] Harshman N L and Ranade K S 2011 {\it Phys. Rev. A} {\bf 84}
012303

[4] Stokes A, Kurcz A,  Spiller P T and Beige A 2012 {\it Phys.
Rev. A} {\bf 85} 053805

[5] Stokes A 2012 {\it  Phys. Rev. A} {\bf 86} 012511

[6] Dugi\' c M and Jekni\' c-Dugi\' c J 2012 {\it Pramana: J.
Phys.} {\bf 79} 199

[7] Arsenijevi\' c M, Jekni\' c-Dugi\' c J and  Dugi\' c M 2013
{\it Chin. Phys. B} {\bf 22} 020302

[8] Dugi\' c M, Arsenijevi\' c M and  Jekni\' c-Dugi\' c J 2013
  {\it Sci. China PMA} {\bf 56} 732

[9] Fel'dman E and Zenchuk A I 2012 {\it Phys.Rev. A} {\bf 86}
012303

[10] Wichterich H C 2011 {\it Entanglement Between
Noncomplementary Parts of Many-Body Systems} (Berlin: Springer
Theses, Springer)

[11] Thirring W, Bertlmann R A, Kohler R and Narnhofer H 2011 {\it
Eur. Phys. J. D} {\bf 64} 181

[12] Harshman N L 2012 arXiv:1210.1770, 2012

[13] Jekni\' c-Dugi\' c J, Arsenijevi\' c M and Dugi\' c M 2013
{\it Quantum Structures: A View of the Quantum World}
(Saarbrucken: Lambert Academic Publishing)

[14] Giulini D, Joos E, Kiefer C, Kupsch J, Stamatescu I-O and Zeh
H D 1996 {\it Decoherence and the Appearance of a Classical World
in Quantum Theory} (Berlin: Springer)

[15] Zurek W H 2003 {\it Rev. Mod. Phys.} {\bf 75} 715

[16] Schlosshauer M 2004 {\it Rev. Mod. Phys.} {\bf 76} 1267

[17] Wallace D 2012 {\it The Emergent Multiverse: Quantum Theory
according to the Everett Interpretation}
 (Oxford: Oxford University Press)

[18] Breuer H P and Petruccione F 2002 {\it The Theory of Open
Quantum Systems} (Oxford: Oxford University Press)

[19] Rivas A. and Huelga S F 2011 {\it Open Quantum Systems. An
Introduction} (Berlin: Springer)

[20] Nakajima S 1958 {\it Prog. Theor. Phys.} {\bf 20} 948

[21] Zwanzig R 1960 {\it J. Chem. Phys.} {\bf 33} 1338

[22] Wheeler A. J. and Zurek W H (Eds.) 1983 {\it Quantum Theory
and Measurement} (Princeton: Princeton University Press)

[23] Gemmer J,  Michel M,  Mahler G 2004 {\it Quantum
Thermodynamics, Lecture Notes in Physics,
  Vol. 657}
  (Berlin: Springer-Verlag)

[24] Scully M. O. and Zubairy M S 2001 {\it Quantum Optics}
(Cambridge: Cambridge University Press)

[25] Nielsen M A and Chuang I L 2000  {\it Quantum Computation and
Quantum Information}   (Cambridge: Cambridge University Press)

[26] Weiss U 1999
  {\it Quantum Dissipative Systems}
  (Singapore: World Scientific)

[27] Pechukas P and Weiss U  (Eds.) 2001
  {\it Quantum Dynamics of Open Systems
  Vol. 268}, Chemical Physics, Special Issue

[28] Narlikar A V and Fu Y Y 2010 {\it The Oxford Handbook on
Nanoscience and Technology: Frontiers and Advances} (Oxford:
Oxford University Press)

[29] Tommasini P, Timmermans E, Piza A F R D 1998 {\it Am. J.
Phys.} {\bf 66} 881

[30] X. Ma,  Zotter S, Kofler J, Ursin R, Jennewein T,  Brukner \v
C and Zeilinger A 2012 {\it Nature Phys.} {\bf 8} 480

[31] Salgado D, Sanches-Gomez J I and Ferrero M 2004 {\it Phys.
Rev. A} {\bf 70} 054102

[32] Tong D M, Kwek L C, Oh C H, Chen J-L and  Ma L 2004 {\it
Phys. Rev. A} {\bf 69} 054102

[33] Shaji A 2005 {\it Ph.D. Thesis}
  (Austin: University of Texas at Austin)

[34] Hayashi H, Kimura G,  Ota Y 2003
  {\it Phys. Rev. A} {\bf 67} 062109

[35] Brodutch A, Datta A, Modi K, Rivas A and Rodríguez-Rosario C
A  2013 {\it Phys. Rev. A} {\bf 87} 042301

\end{document}